# Sedimentation of colloidal plate-sphere mixtures and inference of particle characteristics from stacking sequences

Tobias Eckert, Matthias Schmidt,[*] and Daniel de las Heras[†]

*Theoretische Physik II, Physikalisches Institut, Universität Bayreuth, D-95440 Bayreuth, Germany*

We investigate theoretically the effect of gravity on a plate-sphere colloidal mixture by means of an Onsager-like density functional to describe the bulk, and sedimentation path theory to incorporate gravity. We calculate the stacking diagram of the mixture for two sets of buoyant masses and different values of the sample height. Several stacking sequences appear due to the intricate interplay between gravity, the sample height, and bulk phase separation. These include the experimentally observed floating nematic sequence, which consists of a nematic layer sandwiched between two isotropic layers. The values of the thicknesses of the layers in a complex stacking sequence can be used to obtain microscopic information of the mixture. Using the thicknesses of the layers in the floating nematic sequence we are able to infer the values of the buoyant masses from the colloidal concentrations and vice versa. We also predict new phenomena that can be experimentally tested, such as a nontrivial evolution of the stacking sequence by increasing the sample height in which new layers appear either at the top or at the bottom of the sample.

DOI: 10.1103/PhysRevResearch.4.013189

## I. INTRODUCTION

With remarkable exceptions [1–3], experiments on colloidal science are performed on Earth. Hence, the colloidal particles are subject to a gravitational field that can have a strong effect on the system. The effect of gravity is particularly strong if the suspension contains particles with significantly different buoyant masses, such as e.g., the case of strongly polydisperse and multicomponent colloidal systems. On the other hand, theoretical studies of colloidal systems are often focused on bulk properties and disregard the effect of gravity on the system.

This paper aims at bridging the gap between sedimentation-diffusion-equilibrium experiments and theoretical studies of bulk phenomena in colloidal plate-sphere mixtures. Experimental works include studies on the structure [4] and the rheology [5] of mixtures of silica nanospheres and kaolinite plates, the observation of a slowdown of the crystallization transition of spheres due to the addition of plates [6], the occurrence of isotropic-columnar coexistence in charged mixtures of gibbsite plates and silica spheres [7,8], enhanced density fluctuations of the spheres due to the addition of plates in dilute suspensions [9], the formation of an arrested glass state in gibbsite platelets and silica sphere mixtures [10], and several sedimentation experiments in which a floating nematic layer sandwiched between two isotropic layers [11–13] was observed. From a theoretical point of view, the bulk properties of plate-sphere mixtures have been investigated via free-volume theory [14], density functional theory in the Onsager approximation with [15–17] and without [18] rescaling of the second virial coefficient as well as using fundamental measure density functional theory [15,19–21], with explicit approximations to the configurational partition function [22], and via a density expansion on the work required to insert particles to the mixture [23].

The effect of gravity on a colloidal plate-sphere mixture has received little theoretical attention, with notable exceptions that have analysed the floating nematic stacking sequence [11,24,25]. We use here sedimentation path theory [26] to connect bulk and sedimentation phenomena in colloidal plate-sphere mixtures. The theory is based on the so-called sedimentation paths, which are straight lines in the plane of chemical potentials. The paths represent the linearly varying local chemical potentials along the vertical axis in a sample that is subject to gravity. An interface between two layers of different bulk phases appears in a cuvette if a sedimentation path crosses a bulk binodal. Several stacking sequences can occur by varying the control parameters of the mixture such as the colloidal concentrations and compositions but also as a result of changing the sample height [27,28]. The set of stacking sequences for a given mixture can be grouped in a stacking diagram, which depicts all possible stacking sequences in the plane of experimentally relevant quantities, such as the overall packing fraction for each species. The stacking diagram is in sedimentation-diffusion-equilibrium, the analog of the bulk phase diagram in equilibrium.

[*]Matthias.Schmidt@uni-bayreuth.de
[†]delasheras.daniel@gmail.com; www.danieldelasheras.com

The stacking diagrams of several colloidal mixtures have been calculated with sedimentation path theory using both the infinite sample height limit [26,29–32] and also the case of finite sample height [27–29]. The later allows to carry out a direct comparison with experimental findings. Excellent agreement of results from the sedimentation path theory and the experimental observations by van der Kooij and Lekkerkerker [33,34] has been recently found in mixtures of plates and rods [28].

Here, we use sedimentation path theory to study theoretically the effect of gravity on a colloidal plate-sphere mixture and compare with corresponding sedimentation experiments [11]. In the experiments [11] only isotropic and uniaxial phases were reported. We hence restrict the bulk study to phases without positional order using a simple microscopic density functional theory. The isotropic-nematic bulk binodal has an inflection point in the plane of chemical potentials. The occurrence of an inflection point affects the sedimentation-diffusion equilibrium by enriching the set of possible stacking sequences. We study how the stacking diagram changes by varying both the buoyant masses of the species and the height of the sample. We also demonstrate how to use the macroscopically observed stacking sequences in the experiments to infer microscopic information about the colloidal particles, such as their buoyant masses. Our methodology is general and can be used in other colloidal mixtures to both obtain the stacking diagram and infer particle characteristics from macroscopic stacking behavior.

## II. THEORY

### A. Plate-sphere particle model

Lyotropic liquid crystals are often modelled using hard particles [35] for which the pairwise interparticle potential is infinite if two particles overlap and zero otherwise. We use here a mixture of hard plates and hard spheres to model the experimental colloidal particles of Ref. [11]. In the experiments only isotropic ($I$) and uniaxial nematic ($N$) bulk phases were reported. Hence, we restrict the bulk study to phases without positional order. The nematic phase is rich in the anisotropic particles, i.e., the plates. A schematic of both phases is shown in Fig. 1. The uniaxial order parameters $S_p$ of the plates (see Appendix A) characterizes the isotropic ($S_p = 0$) and the nematic ($S_p > 0$) phases. In what follows, we use subscripts p and s to designate the plates and the spheres, respectively.

The gravitational length of species $i = p, s$ is $\xi_i = k_B T/(m_i g)$ with $m_i$ the buoyant mass of the species, $g$ the gravitational acceleration, $k_B$ Boltzmann's constant, and $T$ absolute temperature. In the experimental study [11], the plates were made of gibbsite (mass density 2.42 g/cm$^3$) and the spheres of alumina-coated silica (2.30 g/cm$^3$). The particles were sterically stabilized with a polymer coating of a few nanometer thickness and suspended in an aqueous solvent (1.00 g/cm$^3$). Here we use cylinders of diameter 184 nm and thickness 2 nm together with spheres of diameter 74 nm to model the cores of the plates and of the spheres, respectively. The core dimensions are relevant to calculate the buoyant masses, and thus the gravitational lengths. The effective dimensions of the coated particles are obtained by adding 10 nm to the core dimensions, as estimated experimentally by neutron scattering [11]. Hence, we use plates of diameter 194 nm and spheres of diameter 84 nm for the hard particle-particle interactions, see Fig. 1. Gibbsite plates are usually polydisperse. For example, in a similar system, the uncertainty in the thickness of the plates is approximately 20% [33,34]. We therefore use the effective plate thickness as an adjustable parameter to match the properties of the isotropic-nematic transition of the monocomponent system of plates between theory and experiments. In the experimental study [11] the packing fractions at the isotropic-nematic coexistence in a pure system of plates are 0.045 and 0.049, respectively. We study the bulk of the mixture with an Onsager-like [36] classical density functional theory [37], see details in Appendix A. We find an effective plate thickness of 2 nm to be the optimal value such that the average between the isotropic and the nematic coexisting densities of a pure system of plates are the same in the theory and in the experiments. The theory however overestimates the density jump at the transition: The predicted coexisting isotropic-nematic packing fractions are 0.042 and 0.052.

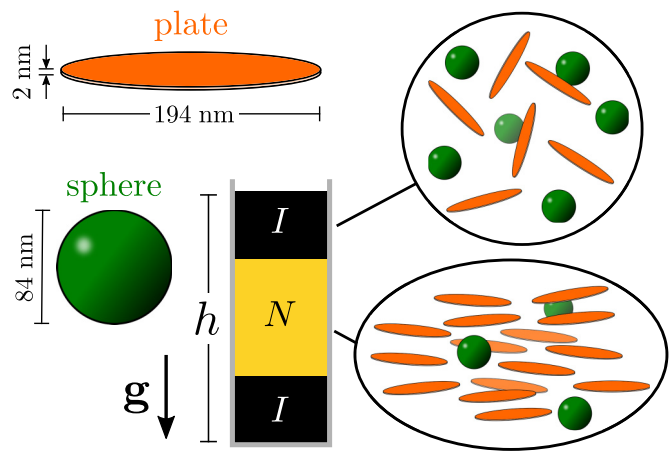

FIG. 1. Dimensions of the hard spheres (green) and hard cylinders (orange) used to model the colloidal plate-sphere mixture, together with a sketch of a cuvette of height $h$ under a gravitational field **g**. The stacking sequence is a floating nematic phase $INI$, i.e., top isotropic, middle nematic, and bottom isotropic. Both the stacking sequence and the thickness of the layers are consistent with one of the experimental samples reported in Ref. [11]. Schematics of the particles in the isotropic (no orientational order $S_p = 0$) and in the uniaxial nematic (orientational order $S_p > 0$) phases are also shown.

We estimate the buoyant masses using only the volumes of the cores. That is, we neglect the effects of the polymer coating since its mass density (1.02 g/cm$^3$) is close to the density of the aqueous solvent. With the above values of the mass densities and particle dimensions of plates and spheres, we obtain the gravitational lengths $\xi_p = 5.34$ mm and $\xi_s = 1.49$ mm. Hence, the buoyant mass ratio in our system is

$$s = \frac{m_s}{m_p} = \frac{\xi_p}{\xi_s} \approx 3.58. \quad (1)$$

Our estimate of the gravitational length of the plates (5.34 mm) is larger than that in Ref. [11] (2.92 mm) due to the adjustable value of the plate thickness. Recall that the effective plate thickness is the only adjustable parameter that we use in our theoretical study. The remaining particle dimensions as well as the mass densities are directly taken as reported in the experimental study [11].

### B. Sedimentation path theory

To incorporate gravity we employ a local density approximation (LDA) that approximates each horizontal slice of the system at height $z$ by a bulk equilibrium system with local chemical potentials $\mu_i(z)$ given by [26,27,29]

$$\mu_i(z) = \bar{\mu}_i - m_i g\left(z - \frac{h}{2}\right), \quad i = \text{p, s}. \quad (2)$$

Here $0 \leqslant z \leqslant h$ is the vertical coordinate measured from the bottom of the sample, $h$ is the height of the sample, $\bar{\mu}_i$ with $i = \text{p, s}$ are the chemical potentials in the absence of gravity, and $m_i g z$ are the gravitational potentials (linear in $z$). The LDA is justified if all correlation lengths are small compared to both gravitational lengths, which is the case in many colloidal systems including the current one. Note that the LDA is used here only to incorporate gravity to the underlying bulk theory. Hence, the LDA does not affect the theoretical treatment of the bulk. The description of the bulk can be done with a simple Onsager theory like we use here, but also with more sophisticated density functional theories [35,38] and other approaches such as perturbation theory [39].

Equation (2) describes a line segment in the plane of chemical potentials. The position (i.e., the statepoint) along the line segment is parameterized by $z$. We refer to such line segments as sedimentation paths [26,27]. Eliminating $z$ for the binary mixture in Eq. (2) yields

$$\mu_s(\mu_p) = s\mu_p + a = s(\mu_p - b), \quad (3)$$

which is the equation of a line segment in the plane of $\mu_p$ and $\mu_s$ with slope given by the buoyant mass ratio $s = m_s/m_p = \xi_p/\xi_s$, intersect $a = \bar{\mu}_s - s\bar{\mu}_p$, and root $b = \bar{\mu}_p - \bar{\mu}_s/s$ [see Fig. 2(a)]. The midpoint of the sedimentation path is $(\bar{\mu}_p, \bar{\mu}_s)$, conveniently translated by the constant terms $m_i g h/2$ in Eq. (2). The length of the path in the plane of $\mu_p$ and $\mu_s$ is $\beta\Delta\mu_i = h/\xi_i$, with $\Delta\mu_i = \mu_i(0) - \mu_i(h)$ and $\beta = 1/(k_B T)$.

The significance of the sedimentation path is that whenever a path crosses a bulk binodal, an interface between the two bulk phases that coexist at the binodal appears in the cuvette, see Fig. 2(a). The crossings between the sedimentation path and the binodal provide therefore the sequence of layers in the sample, i.e., the stacking sequence. Moreover, the value of the parameter $z$ at the crossing dictates the vertical position of the interface in the sample.

### C. Stacking diagram

Depending on the position, the slope, the length, and the direction of the sedimentation path, different stacking sequences can occur. The stacking sequences can be grouped in a stacking diagram. Similar to the bulk phase diagram, the stacking diagram admits several representations that differ in the variables that are kept constant. To compare with experiments we fix the buoyant mass ratio $s$ and the path length, i.e., we fix the buoyant masses of both species and the sample height $h$.

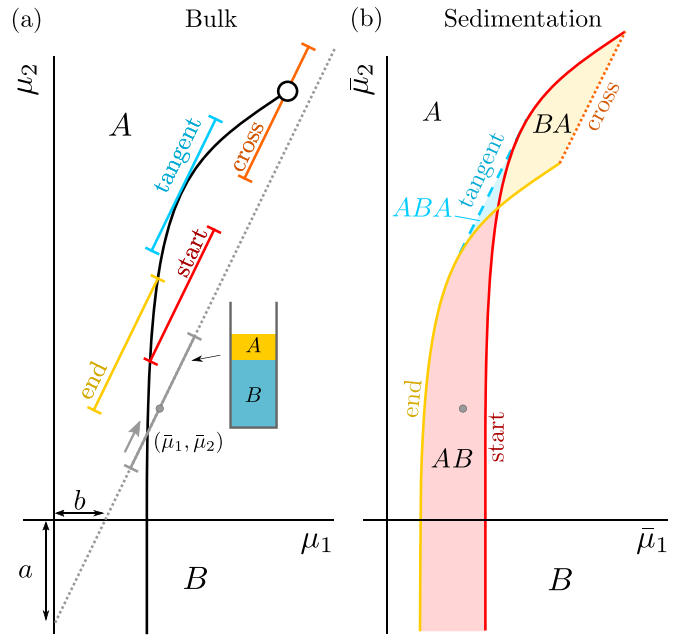

FIG. 2. (a) Model bulk phase diagram in the plane of chemical potentials $\mu_1$ and $\mu_2$. The phases $A$ and $B$ coexist along a binodal (black-solid line) that ends at a critical point (empty circle). The line segments are finite sedimentation paths. The gray path crosses the binodal and corresponds to a stacking sequence $AB$ (from top to bottom). The grey arrow indicates the direction of all paths from top to bottom. Illustrative examples of paths that form boundaries between different stacking sequences are depicted: (i) paths that start (red) or end (yellow) at the binodal, (ii) paths tangent (blue) to the binodal, and (iii) paths that cross (orange) the critical point. A displacement of any of such paths can alter the stacking sequence. The dotted-gray line is a sedimentation path in the limit of infinite height ($a$ and $b$ are the intersects of the path with the $\mu_2$ and the $\mu_1$ axes, respectively). (b) Stacking diagram, plane of average chemical potentials $\bar{\mu}_1$ and $\bar{\mu}_2$, of the bulk phase diagram depicted in (a). Each region is a different stacking sequence, as indicated. The boundary lines between sequences are sedimentation binodals of type I (solid lines) or type II (dashed-blue line), and a terminal line (dotted-orange line). A sedimentation path in (a) is a point in (b) given by the coordinates of the average chemical potentials along the path. See, e.g., the grey circle in (b) that corresponds to the gray sedimentation path (finite height) in (a).

To illustrate the construction of a stacking diagram, we plot in Fig. 2 a hypothetical bulk diagram and its corresponding stacking diagram. In bulk, two phases $A$ and $B$ coexist along a binodal that ends at a critical point, Fig. 2(a). We construct the stacking diagram by finding the sedimentation paths in the bulk phase diagram that form the boundaries between two stacking sequences in the stacking diagram. There exist three types of boundaries [27,28]. The first type, so-called sedimentation binodals of type I, corresponds to sedimentation paths that either start or end at a bulk binodal [red and yellow paths in Fig. 2(a)]. Paths that cross an ending point of a binodal, e.g., a triple point or a critical point [orange path in Fig. 2(a)], also form a boundary in the stacking diagram, known as a terminal line. The third type, known as sedimentation binodal of type II, corresponds to paths tangent to a binodal, see the blue path in Fig. 2(a). An infinitesimal displacement of the midpoint $(\bar{\mu}_1, \bar{\mu}_2)$ of the path can alter the stacking sequence in each of the three cases.

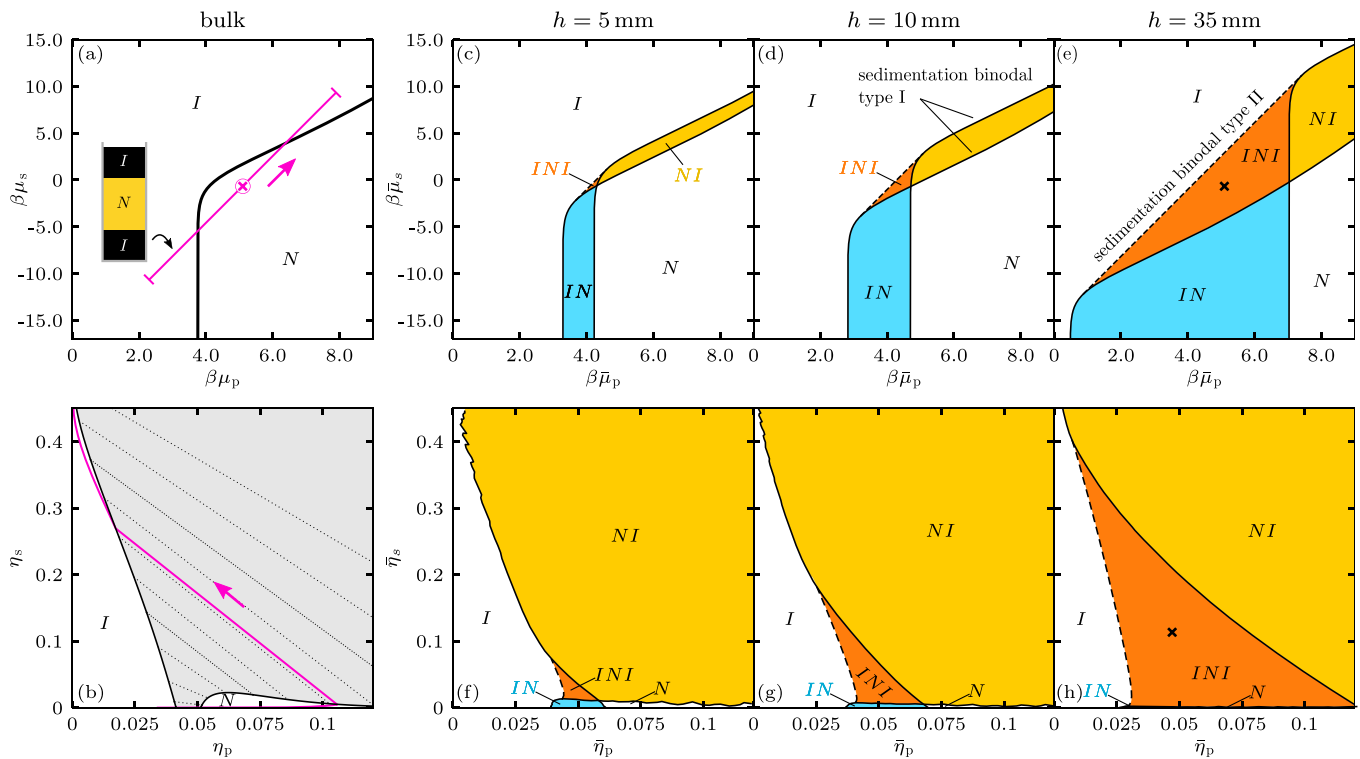

FIG. 3. Bulk phase diagram in the plane of chemical potential of plates $\mu_p$ and spheres $\mu_s$ (a), and also in the plane of packing fractions of plates $\eta_p$ and spheres $\eta_s$ (b). The solid-black line in (a) is the binodal at which the isotropic $I$ and the nematic $N$ phases coexist. Dotted lines in (b) are tie lines connecting coexisting points along the binodal (solid-black line). The grey area is the two-phase region. The pink line in (a) is a sedimentation path corresponding to the $INI$ stacking sequence with layer thicknesses of 9.3 mm, 16.5 mm, and 9.6 mm as found experimentally in Ref. [11]. The pink curve in (b) is the same sedimentation path in the plane of packing fractions. The pink arrows indicate the direction of the path from the top to the bottom of the sample. Stacking diagram in the plane of average chemical potential of plates $\bar{\mu}_p$ and spheres $\bar{\mu}_s$ [(c)–(e)], and in the plane of average packing fractions of plates $\bar{\eta}_p$ and spheres $\bar{\eta}_s$ [(f)–(h)] for three different sample heights: $h = 5$ mm [(c),(f)], 10 mm [(d),(g)], and 30 mm [(e),(h)]. Each colored region correspond to a different stacking sequence (except the pure sequences $I$ and $N$ depicted in white). The sequences are labeled from the top to the bottom of the sample. The black crosses in panels (e) and (h) indicate the position of the sedimentation path plotted in panels (a) and (b). In the stacking diagram, sedimentation binodals of type I (type II) are represented with solid (dashed)-black lines.

The coordinates $(\bar{\mu}_1, \bar{\mu}_2)$ for each of these special paths are then represented in a diagram to produce the stacking diagram in the plane of $\bar{\mu}_1$ and $\bar{\mu}_2$, see Fig. 2(b). However, in the experiments the relevant quantities are usually the colloidal packing fractions. Hence, in our plate-sphere mixture we transform the stacking diagram in the plane of $\bar{\mu}_p$ and $\bar{\mu}_s$ by averaging the packing fraction of each species, i.e., $\eta_i = \rho_i v_i$, along the sedimentation path to obtain $(\bar{\eta}_p, \bar{\eta}_s)$. Here $v_i$ is the particle volume of species $i$. For this transformation we need to compute $\eta_i$ for each point $(\mu_p, \mu_s)$ along the path (see [Appendix A]). In both planes, the $\bar{\mu}_p\bar{\mu}_s$-plane and the $\bar{\eta}_p\bar{\eta}_s$-plane, each point of the stacking diagram represents a sedimentation path, i.e., one sample in sedimentation-diffusion-equilibrium.

## III. RESULTS

### A. Bulk and stacking diagrams

The bulk phase diagram according to our microscopic density functional theory is shown in Figs. 3(a) and 3(b) in the planes of chemical potentials and packing fractions, respectively. We restrict the study to phases without positional order. Two phases occur: isotropic ($I$) with no orientational order of the plates ($S_p = 0$) and uniaxial nematic ($N$) with plates aligned on average along the director ($S_p > 0$). The mixture does not show any critical behavior, nor does it show isotropic-isotropic demixing. Critical and triple points are therefore not present in the bulk phase diagram. The curvature of the binodal in the plane of chemical potentials changes, leading to an inflection point. We see below how the inflection point affects the sedimentation-diffusion-equilibrium of the mixture.

A sedimentation path ($h = 35$ mm and $s = 3.58$) corresponding to a floating nematic stacking sequence $INI$ is depicted in the chemical potential representation of the bulk diagram, see Fig. 3(a). To illustrate the advantage of using the chemical potentials in the description of sedimentation-diffusion equilibrium, we also depict the path in the plane of packing fractions, see Fig. 3(b). The simple form of the path in the plane of chemical potentials (line segments) is lost in other representations of the bulk diagram. In the plane of $\bar{\eta}_p$ and $\bar{\eta}_s$ the path discontinuously jumps from the isotropic to the nematic phase, and vice-versa, along the tie lines that connect coexisting points in bulk. For a detailed study of sedimentation paths in the plane of packing fractions see Ref. [31].

The finite height stacking diagram in the plane of chemical potentials and also in the plane of packing fractions is shown in Figs. 3(c)–3(e) and Figs. 3(f)–3(h), respectively, for three sample heights: 5 mm, 10 mm, and 35 mm. In all cases the buoyant mass ratio is fixed to $s = 3.58$. There are five different stacking sequences, namely $I$, $N$, $IN$, $NI$, and $INI$. We label the sequences from the top to the bottom of the sample. For example, $NI$ indicates a top nematic layer and a bottom isotropic layer. The occurring sequences are all possible ordered subsets of the sequence $INI$. We can arrive at the sequence $I$ by removing from the sequence $INI$ the two top layers, the two bottom layers, or the middle $N$ layer. Since no isotropic-isotropic demixing is observed, the resulting sequence ($I$) through any of these routes is the same. Note that the observation of two isotropic layers in the $INI$ sequence does not imply the occurrence of isotropic-isotropic demixing in bulk.

There is no topological change by varying the sample height from 5 mm to 35 mm. However, the region occupied by a given stacking sequence changes with $h$. Hence, the stacking sequence of two sedimentation-diffusion-equilibrium samples that share either the same ($\bar{\mu}_p, \bar{\mu}_s$) or the same ($\bar{\eta}_p, \bar{\eta}_s$) can change with the sample height $h$. All regions corresponding to stacking sequences with multiple layers grow in the plane of $\bar{\mu}_p$ and $\bar{\mu}_s$ with increasing $h$. In contrast, in the plane of $\bar{\eta}_p$ and $\bar{\eta}_s$ only the $INI$ region grows in size. All the other regions shrink with increasing $h$. In the limit of $h \to \infty$ only the stacking sequences $I$ and $INI$ remain. This is confirmed by the calculation of the stacking diagram in the infinite height limit, see Sec. III E. That only two sequences remain if $h \to \infty$ can be also concluded from the bulk diagram in Fig. 3(a) by lengthening the depicted sedimentation path. The resulting path either lies entirely in the $I$ phase or it transitions into the $N$ phases and hence cuts the binodal twice giving rise to the $INI$ sequence (note that the binodal does not end at a critical point and that the path is an infinite line in the limit $h \to \infty$).

### B. Reading microscopic information from experimental photographs

A prominent result of sedimentation-diffusion-equilibrium experiments is the stacking sequence. Sometimes, the sequence can be easily read from direct visual inspection of the sample using crossed polarizers. Between crossed polarizers, isotropic layers appear dark whereas layers in which the particles posses orientational order are bright. Even two layers of different phases with orientational order, such as, e.g., nematic and columnar, can be differentiated by their relative brightness and color [34,40]. We explore here the possibility of obtaining microscopic information about the particles by using from the experiments only the stacking sequence and the thicknesses of the occurring layers.

The thicknesses of the layers in stacking sequences provide information to locate the corresponding sedimentation path in the plane of chemical potentials. To contain sufficient amount of information and hence be useful for the analysis, such sequences need to possess at least three layers. That is, at least two crossings between the corresponding path and the bulk binodal(s) are required. For example, in Fig. 3(a) we construct the sedimentation path such that its slope is $s = 3.58$ and its stacking sequence is $INI$ with layer thicknesses of 9.3 mm (bottom isotropic), 16.5 mm (middle nematic), and 9.6 mm (top isotropic). The values of the thicknesses are chosen to reproduce the experimental sample in Ref. [11] with 74 nm spheres and average packing fractions $(\bar{\eta}_s, \bar{\eta}_p) = (0.05, 0.05)$.

A path is defined by four values and its direction (given by the sign of the buoyant mass of one species). For example, a path is defined by the position of the two endpoints in the plane of chemical potentials, or by the set of variables $s$, $\Delta\mu_p$, $\bar{\mu}_p$, and $\bar{\mu}_s$. The slope and the value of the three layer thicknesses in the $INI$ sequence give in total four constrains, and hence properly define a unique sedimentation path, see Fig. 3(a).

To locate the path in the previous example, we used fixed values of the buoyant masses. The buoyant masses determine the slope and, together with the sample height, the length of the sedimentation path. Determining experimentally the buoyant masses might be a difficult task since it requires detailed measurements of the particle dimensions and mass densities. We show next that using as input only the experimentally reported thicknesses of the layers in the $INI$ stacking sequence, one can infer a range of gravitational lengths (and also a range of average colloidal concentrations) in which such sample can exists. We therefore pretend that the gravitational lengths are unknown, and consider a wide range of candidate values for both $\xi_p$ and $\xi_s$. For each pair of $\xi_p$ and $\xi_s$, we find the sedimentation path that produces the $INI$ sequence with layer thicknesses equal to those in the experiments, see sketch in the inset of Fig. 4(a). The gravitational lengths fix the length and the slope of the path. Hence, we only vary the position of the path $(\bar{\mu}_p, \bar{\mu}_s)$ until the correct layer thicknesses are reproduced. If a solution exists, the path is unique.

The results are summarized in Fig. 4. Each point corresponds to a path with the sequence $INI$ and layer thicknesses 9.3 mm, 16.5 mm, and 9.6 mm. Illustrative paths of different slope and length in the plane of $\mu_p$ and $\mu_s$ are shown in the inset of Fig. 4(a). All the paths give rise to the desired $INI$ stacking sequence with the correct layer thicknesses. Note that paths of different lengths in the plane of chemical potentials can represent samples with the same height if the paths have different gravitational lengths since $h = \beta\Delta\mu_i\xi_i$.

Once a sedimentation path with the right sequence and layer thicknesses is found, we calculate the corresponding overall packing fractions by integrating the local packing fraction along the path. Figures 4(a) and 4(b) show for each pair $(\xi_p, \xi_s)$ the corresponding values of $\bar{\eta}_p$ and $\bar{\eta}_s$ (see color bars). Above the red-dotted line it is not possible to find an $INI$ sequence because the paths there are too flat (small slope) to cross the bulk binodal twice. From the slope of the bulk binodal at its inflection point, we determine the minimum slope for an $INI$ sequence to occur to be $s_{\min} \approx 1.7$. For slopes $s > s_{\min}$ it is always possible to find an $INI$ sequence, but

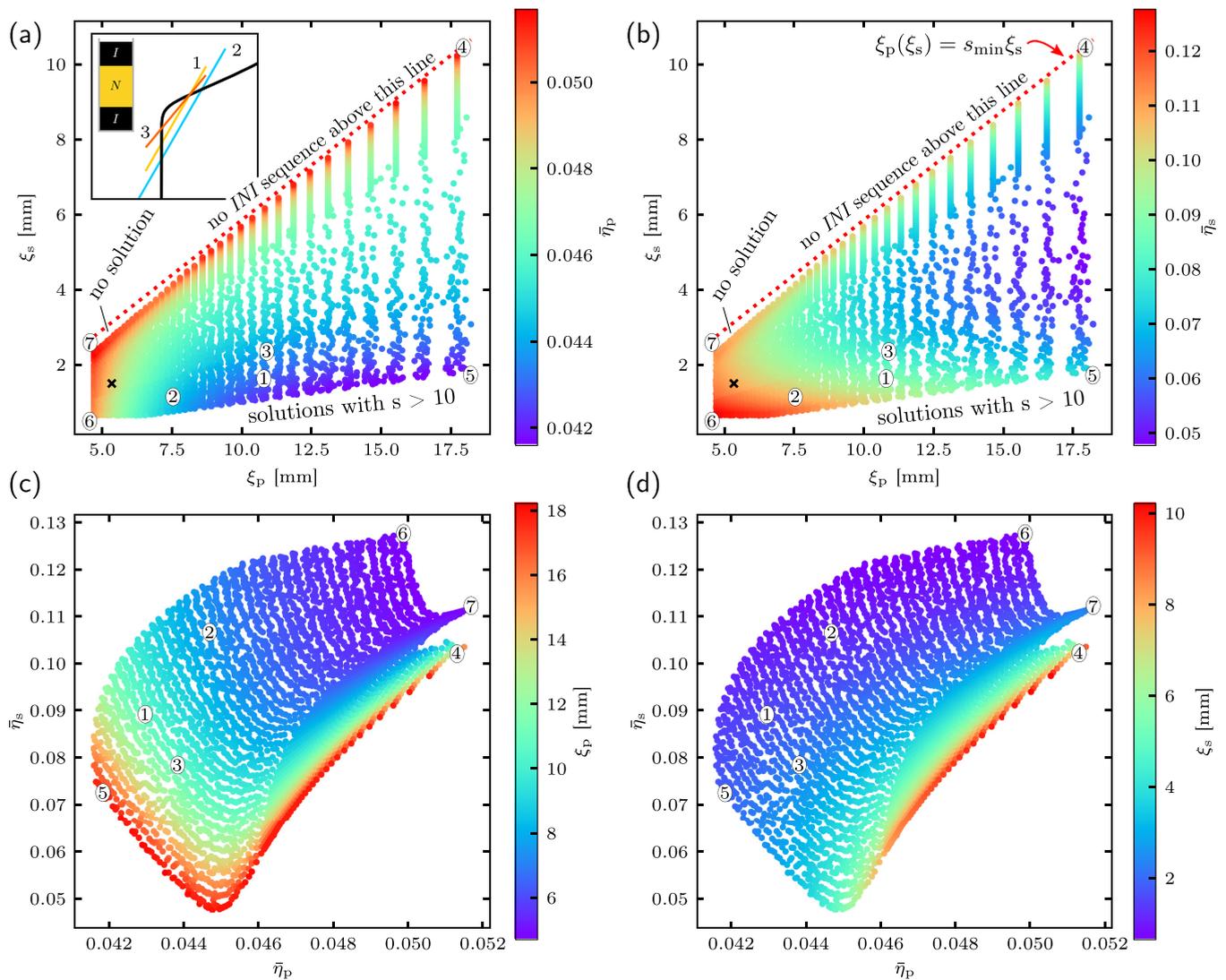

FIG. 4. Average packing fraction of plates $\bar{\eta}_p$ (a) and spheres $\bar{\eta}_s$ (b) indicated by the color map as a function of the gravitational lengths of plates $\xi_p$ and spheres $\xi_s$. Each point in the diagrams represents a sedimentation path that produces an isotropic-nematic-isotropic (*INI*) stacking sequence with layer thicknesses of 9.3 mm, 16.5 mm, and 9.6 mm, from top to bottom. Illustrative paths together with a sketch of the sample are shown in the inset of panel (a). No solution exists above the red-dotted line in panels (a) and (b) due to the path being too flat to cross the binodal twice. The inverse relations, i.e., gravitational lengths as a function of average packing fractions, are depicted in panels (c) and (d). The samples labeled 1–3 correspond to the paths shown in the inset of panel (a). The samples labeled 4–7 are close to the boundaries of the diagrams shown in panels (a) and (b). The black crosses in (a) and (b) indicate the value of the gravitational lengths used here.

there is a small region in which there is no *INI* sequence with the desired thicknesses (region indicated as "no solution" in Fig. 4). For the remaining values of gravitational lengths an *INI* sequence with the desired thicknesses always exists. In Fig. 4 we show only solutions for $s < 10$, which covers a vast range of experimentally realisable buoyant mass ratios. For $s > 10$ the path is almost parallel to the vertical region of the binodal, which makes it difficult to numerically find the solution. The region in which solutions with $s > 10$ can occur is indicated in Figs. 4(a) and 4(b)

The inverse relation to Figs. 4(a) and 4(b) is shown in Figs. 4(c) and 4(d). There, for each pair $(\bar{\eta}_p, \bar{\eta}_s)$ the gravitational lengths, $\xi_p$ and $\xi_s$, required to obtain the correct layer heights are depicted via a color map. This representation clearly shows that given a specific set of layer thicknesses, not every combination of packing fractions is possible. Also, the set of possible values of $\bar{\eta}_p$ is limited to the narrow range [0.0415, 0.052]. The possible concentration of spheres varies in a wider range of approximately [0.05, 0.13]. We expect these ranges to slightly increase if solutions with $s > 10$ are also considered.

The four-sided shape of the point cloud in the plane of $\xi_p$ and $\xi_s$ in Fig. 4(a) is deformed when transformed into the plane of $\bar{\eta}_p$ and $\bar{\eta}_s$ in Fig. 4(c). A nonlinear function governs this transformation from gravitational lengths to average packing fractions. Despite the nonlinearity, the clockwise order of the points labeled 4–7 is preserved under this transformation.

Using the experimental values for the layer thicknesses in the *INI* sequence together with the values of the gravitational lengths known from the synthesis of the particles, we can use Figs. 4(a) and 4(b) to infer the values of the packing fractions. The opposite also holds: If the packing fractions are known from the preparation of the samples, but the microscopic gravitational lengths are unknown, we can use Figs. 4(c) and 4(d) to infer their values.

For the gravitational lengths calculated with our particle model, $\xi_p = 5.34$ mm and $\xi_s = 1.49$ mm [marked in Figs. 4(a) and 4(b) with black crosses], we predict using Figs. 4(a) and 4(b) that the average packing fractions of the sample are $\bar{\eta}_p = 0.047$ and $\bar{\eta}_s = 0.11$. This is in almost perfect agreement for the plates (the packing fraction used in the experimental study [11] is $\bar{\eta}_p = 0.05$). For the spheres, our predicted concentration differs by a factor of two (in the experimental study [11] $\bar{\eta}_s = 0.05$). The difference might be due to our simple theoretical description of the bulk and also to intrinsic characteristics of the experiments, such as polydispersity and the uncertainties in the dimensions of both the particles' cores and the thickness of the coating polymer layer. Note that our discrepancy of a factor of two in the packing fraction of the spheres can be explained with a change in the diameter of the spheres of approximately $\sqrt[3]{2} \approx 1.26$. This value is compatible with the variance in the distribution of the diameter of the spheres (26%) due to polydispersity [9].

Using the stacking diagram, e.g., Figs. 3(g)–3(i), we can choose the packing fractions that produce a given stacking sequence for fixed buoyant mass ratio $s$. The type of analysis done in Fig. 4 allows us to choose the parameters that produce not only the sequence but also the desired layer thicknesses within the sequence.

### C. Effects of a bulk inflection point on the stacking diagram

The Onsager-like density functional theory used here predicts that in the plane of chemical potentials the isotropic-nematic bulk binodal presents an inflection point. The fundamental measure theory applied to a mixture of infinitely thin plates and spheres [15,20] also predicts the occurrence of an inflection point [11], which therefore seems to be a robust feature of the system.

The inflection point does not have any qualitative effects on the stacking diagram for a slope of $s = 3.58$. However, for slopes of the path comparable to the slope of the bulk binodal at the inflection point, the inflection point induces topological changes to the stacking diagram due to the occurrence of new sequences.

The maximum number of layers that can appear in a stacking sequence is [26]

$$l_{\max} = 3 + 2(n_b - 1) + n_i, \quad (4)$$

with $n_b$ the number of bulk binodals and $n_i$ the total number of inflection points in all the binodals. The occurrence of several layers in sedimentation-diffusion-equilibrium is unrelated to bulk coexistence in which the Gibbs phase rule dictates the maximum number of phases that can coexist simultaneously (with the notable exceptions found recently in colloid-polymer mixtures [41–43] in which by fine-tuning the interparticle interactions it is possible to find bulk multiphase coexistence involving more than three different phases). Under gravity, the maximum number of layers in a sequence, Eq. (4), is achieved if a path crosses each binodal for the maximum number of possible times (i.e., two plus the number of inflection points of the binodal). Whether or not a sequence with $l_{\max}$ layers can actually occur depends on the position of the binodals relative to each other in bulk. Here $n_b = 1$ and $n_i = 1$, which yields $l_{\max} = 4$. To confirm the general argument presented above, we see below that in our mixture the sequence with $l_{\max}$ layers is *ININ*. This sequence occurs in the range of buoyant mass ratios $s \in [1.7, 2.5]$ provided that the path is long enough. The slope must be larger than the slope of the binodal at the inflection point ($s \gtrsim 1.7$) but also smaller than the slope of the binodal in the limit as both chemical potentials approach infinity ($s \lesssim 2.5$).

We choose a buoyant mass ratio of $s = 2$, i.e., slightly above the slope of the binodal at the inflection point. The stacking diagram in the plane of average chemical potentials and average packing fractions for sample heights 20 mm, 32 mm, and 40 mm is shown in Fig. 5. A prominent feature is the presence of the four layer stacking sequence *ININ* [11,24] for the sample heights 32 mm and 40 mm. This sequence can only occur for sufficiently large samples ($h \gtrsim 30$ mm) since the sedimentation path needs to start in the *I* phase, enter the *N*, reenter the *I* phase again, and finally end in the *N* phase. For $h = 20$ mm the path is not long enough and it can either start in the *I* phase but not reach the ultimate *N* phase (giving rise to the sequence *INI*) or end in the *N* phase without having started in the *I* phase (resulting in *NIN*).

Hence, from $h = 20$ mm to $h = 32$ mm a topological change in the stacking diagram occurs. The type I sedimentation binodal that separates the sequences *NI* from *INI*, and the type I sedimentation binodal that separates *NI* from *NIN* cross each other, c.f. panels (d) and (e) in Fig. 5. This gives rise to the *ININ* sequence. The same two sedimentation binodals no longer cross for $h = 40$ mm, which eliminates the sequence *NI* entirely in favor of *ININ*, cf. panels (e) and (f) in Fig. 5. Both topological changes can also be observed in the plane of $\bar{\mu}_p$ and $\bar{\mu}_s$. For $h = 20$ mm the two type I sedimentation binodals (same shape as the bulk binodal) intersect each other twice due to the inflection point, see Fig. 5(a). For $h = 32$ mm one of the intersection points has moved over to the other side of one of the points of tangency, see inset of Fig. 5(b), and the *ININ* stacking sequence appears. The two intersections between the two sedimentation binodals of type I merge approximately for $h = 40$ mm into a single point, see Fig. 5(c). As a consequence the *NI* stacking sequence disappears. The transition from $h = 20$ mm to $h = 40$ mm replaces therefore the sequence *IN* by *ININ* but leaves the other sequences unaltered (except for changes in the shape of the regions occupied by each sequence that are inherent to changes in the height).

Another notable feature of the stacking diagram is the presence of a reentrant *IN* stacking sequence. This sequence (blue in Fig. 5) appears in significantly different and disconnected regions of the stacking diagram corresponding to low and high average packing fractions of colloids. The two *IN* regions share in common a direct connection to both the *I* and the *N* regions. This is clearly visible in the chemical potential representation of the stacking diagram, see Figs. 5(a)–5(c). Reentrant phenomena can occur in the bulk of mono- [44,45] and multicomponent [46,47] systems, but also, as in the present case, induced by external fields [48–50].

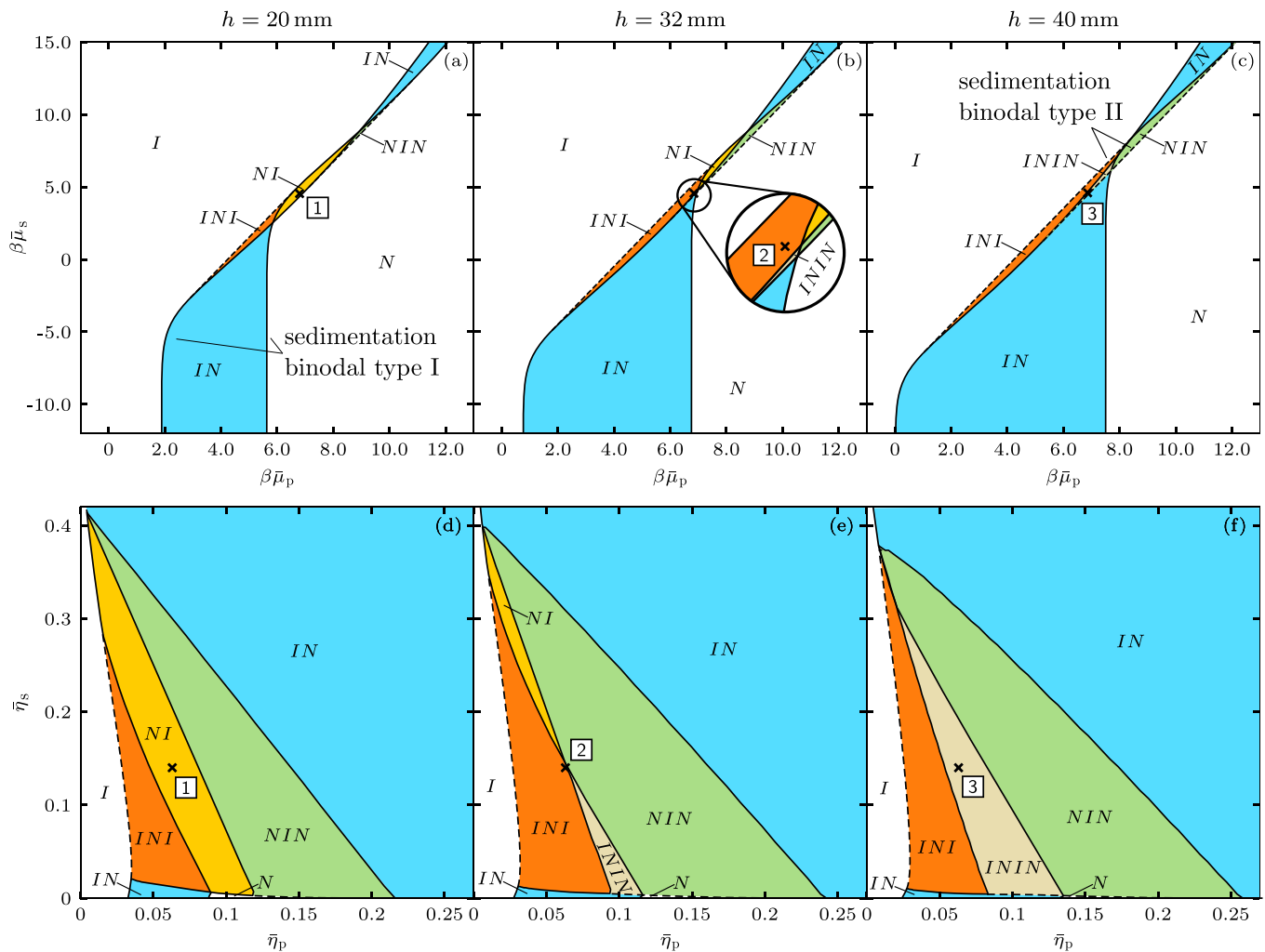

FIG. 5. Stacking diagram in the plane of average chemical potential of plates $\bar{\mu}_p$ and spheres $\bar{\mu}_s$ [(a)–(c)] and also in the plane of average packing fractions of plates $\bar{\eta}_p$ and spheres $\bar{\eta}_s$ [(d)–(f)] for sample heights of $h = 20$ mm [(a), (d)], 32 mm [(b), (e)], and 40 mm [(c), (f)]. The buoyant mass ratio is $s = 2$ in all cases. The crosses indicate at each height the location of the samples with average packing fractions $(\bar{\eta}_p, \bar{\eta}_s) = (0.063, 0.14)$ labeled with white squares as samples 1, 2, and 3. The stacking sequences are labeled from top to bottom of the sample. The inset in (b) is a close view of a small region of the stacking diagram. Sedimentation binodals of type I (type II) are represented with solid (dashed)-black lines.

### D. Dependency on sample height

The stacking diagram for buoyant mass ratio $s = 2$ shows a strong dependency on the sample height. To further investigate the influence of the sample height on the stacking sequence, we consider a set of samples with the same average colloidal packing fractions, $(\bar{\eta}_p, \bar{\eta}_s) = (0.063, 0.14)$, but different heights. This corresponds to an experimental setup with a solution prepared with the desired concentrations, which is then distributed into cuvettes filled to a different height. Only a single colloidal solution needs to be prepared, and even if there is a large uncertainty in the colloidal concentrations, it is at least guaranteed that the concentrations are the same throughout all cuvettes.

Here, we compute for each height $h$ in a range from 2 mm to 42 mm the stable phase ($I$ or $N$) that would be observed in an experiment as a function of the elevation $z$. We then plot the results in the plane of $z$ and $h$, see Fig. 6. This is a different representation of the stacking diagram in which the colloidal concentrations are kept constant. Samples from 2 mm to 30 mm always show the same $NI$ stacking sequence. At 30 mm an additional isotropic layer evolves at the top of the sample, followed by the emergence of a nematic layer at the bottom of the sample from 38 mm onwards. In total two additional layers form, one at the bottom ($N$) and one at the top ($I$) of the sample, as compared to the initial $NI$ sequence for low heights.

### E. Stacking diagram for samples with infinite height

So far, we have studied the stacking diagram for finite height samples and for two fixed values of the buoyant mass ratio, $s = 3.58$ and $s = 2$. We end the results section showing that these two illustrative values of the buoyant mass ratio give rise to the stacking diagrams with the two largest possible number of stacking sequences. To this end, we use sedimentation path theory for samples with infinite height [26,29]. In the limit of infinite height, a sedimentation path is a straight line (not a segment as in the finite case) that can therefore be described using only the slope $s$, the root $b$ [see Eq. (3) and Fig. 2] and the direction of path.

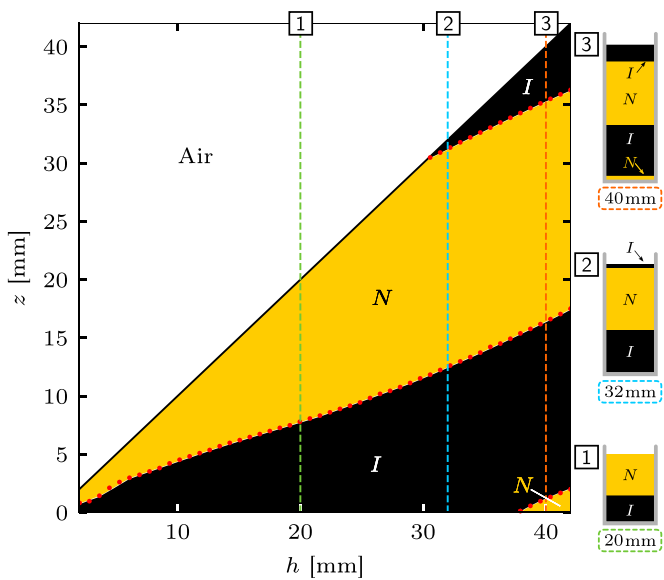

FIG. 6. Stacking diagram at fixed colloidal concentrations. Stable layer at elevation $z$ as a function of the total sample height $h$ for samples with fixed concentrations $(\bar{\eta}_p, \bar{\eta}_s) = (0.063, 0.14)$. The buoyant mass ratio is $s = 2$. The vertical-dashed lines mark the samples 1, 2, and 3 with heights 20 mm, 32 mm, and 40 mm. The same samples are also labeled and marked with crosses in Fig. 5. The solid-black line indicates the sample-air interface.

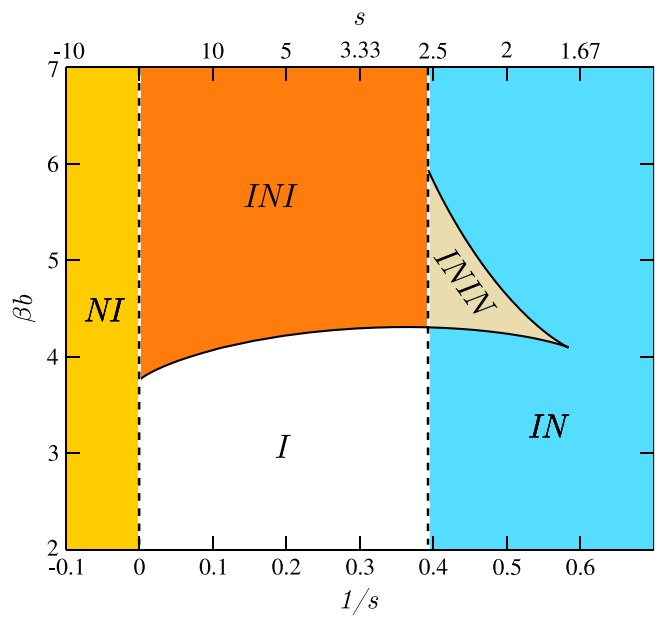

FIG. 7. Stacking diagram for samples in the limit of infinite height in the plane of inverse slope $1/s$ and root $b$ of the sedimentation path. The auxiliary-horizontal axis indicates the value of the slope $s$. Each region corresponds to a different stacking sequence in the limit of infinite height. Sequences are labeled from top to bottom of the sample if $m_p > 0$ and from bottom to top if $m_p < 0$. Black-solid lines are sedimentation binodals formed by the paths that are tangent to the bulk binodal. Vertical-dashed lines are asymptotic terminal lines formed by the two sets of paths that are parallel to the bulk binodal in the limits $\mu_s \to -\infty$ ($1/s = 0$) and $\mu_s \to +\infty$ ($1/s \approx 0.39$).

Similarly to the finite height case, there exist special paths that are boundaries between two different stacking sequences [26,29]: paths tangent to a binodal, paths crossing an ending point of a binodal, and paths parallel to the asymptotic behavior of the binodal at $\mu_i \to \pm\infty$. Plotting the coordinates of the special paths results in a stacking diagram for infinite height, which gives a global overview of sedimentation phenomena for all possible buoyant mass ratios.

For the plate-sphere mixture considered here, the stacking diagram for infinite height in the plane of $1/s$ and $\beta b$ is shown in Fig. 7. Note that we use $1/s$ to represent the slope, since paths parallel to the binodal in the limit $\mu_s \to -\infty$ are vertical, i.e., $s \to \infty$. For convenience, we also indicate the slope $s$ in the auxiliary $x$ axis of Fig. 7.

From $s \approx 1.7$ to $s \approx 2.5$ we find the four layer stacking sequence $ININ$, which we investigated in detail for the illustrative slope $s = 2$ in Fig. 5. For all negative slopes ($s < 0$) the sequence $NI$ is the only possible stacking sequence. Positive slopes up to $s \approx 1.7$ exclusively produce the $IN$ stacking sequence. For $s \gtrsim 2.5$ we find the stacking sequences $INI$ and $I$, depending on the root of the sedimentation path.

The stacking sequences that occur in finite samples are always subsequences of the infinite height limit. Hence, we conclude from Fig. 7 that $s = 2$ and $s = 3.58$ capture the essence and all the interesting phenomenology of the binary mixture of plates and spheres. This includes, among others, floating nematic phases $INI$, a four-layer stacking sequence $ININ$, and the occurrence of an $IN$ reentrant sequence.

## IV. CONCLUSIONS

We have studied the sedimentation-diffusion-equilibrium of a simple binary mixture with only one anisotropic species and restricting the bulk to isotropic and uniaxial nematic phases (i.e., no positional order). The corresponding stacking diagram is substantially richer than the bulk phase diagram, with stacking sequences made of up to four layers and the occurrence of reentrant sequences. The topology of the stacking diagram depends on the sample height and on the buoyant mass ratio. An analysis of the stacking diagram in the limit of samples with infinite height reveals that the two buoyant mass ratios considered here, $s = 2$ and $s = 3.58$, capture most of the sedimentation phenomenology of the mixture. Experimentally, it might be possible to alter the buoyant mass ratio by changing the material of the colloidal cores and/or the solvent mass density. Alternatively, using magnetic colloidal spheres [51,52] and an external magnetic field parallel to the gravitational field should effectively have the same effect as varying the buoyant mass ratio of the mixture.

Analysing the effect of the sample height, we have seen that layers appear in a stacking sequence at both the top and the bottom of the sample by increasing the height while keeping the colloidal concentrations constant. Layers can also disappear from the sequence in the middle of a sample [28] such that formerly separated layers merge by changing the sample height. The vanishing of a middle layer has not been observed here for the considered heights and buoyant mass ratios. The phenomenology found here is therefore complementary to the observation in Ref. [28] and highlights again the relevance of the sample height in sedimentation experiments [27]. Moreover, it shows that *a priori* one does not know where the next layer will form or vanish when transitioning from one sequence to another.

Stacking sequences with three or more layers are often found in sedimentation experiments on colloidal mixtures [11,33,34,53,54]. If both the buoyant masses and the bulk of the mixture are known, the value of the thicknesses of the layers in one of such multi layer sample is enough to uniquely locate the corresponding sedimentation path [11,28]. Knowledge of the particle dimensions is required to construct a theory for the bulk. We have shown here that it is also possible to use the layer thicknesses and the bulk diagram to find the set of all possible sedimentation paths associated to the sample. The set of paths can then be used to estimate the buoyant masses via the colloidal concentrations and vice versa. Including the dimensions of the particles as additional free parameters (i.e., allowing the bulk behavior of the mixture to change) is also possible. Then, a set of experimental samples that differ in their sample heights could provide sufficient information to estimate microscopic parameters such as the particle dimensions and buoyant masses. The required information from the experiments would simply consist of the layer thicknesses, which might be directly measured from the sample images.

The hard particle models used here for both the spheres and the plates are monodisperse. However, size- and therefore mass-polydispersity are inherent to essentially any colloidal system. Several works have considered the effect of polydispersity on the bulk phenomena of a system, see, e.g., Ref. [55] for a review. In contrast, very little is known about the interplay between polydispersity and gravity in sedimentation-diffusion-equilibrium. We will report on the extension of sedimentation path theory to polydisperse colloidal systems in a future publication.

Sedimentation path theory is based on a local equilibrium condition. A sample under gravity is described as a collection of bulk systems with local chemical potentials fixed according to the value of the vertical coordinate. The theory can be used to describe sedimentation in any colloidal mixture, including polymer-colloid mixtures. The addition of polymers to a colloidal suspension can be used to tune the bulk phase behavior of the colloids [56,57] and therefore also the stacking diagram.

An approach conceptually similar to sedimentation path theory that also relies on local equilibrium conditions has been recently used to study sedimentation profiles of molecular systems in centrifugal fields [58]. Even though the gravitational field is not constant under centrifugation, the sedimentation paths are still lines in the space of chemical potentials. Hence, following the ideas of sedimentation path theory it should be possible to construct the stacking diagrams of both colloidal [59] and molecular mixtures under centrifugation.

## ACKNOWLEDGMENTS

This work is supported by the German Research Foundation (DFG) via Project No. 436306241. Publication costs partially funded by the DFG Project No. 491183248 and the Open Access Publishing Fund of the University of Bayreuth.

## APPENDIX: METHODS

**Bulk phase behaviour**. We use classical density functional theory (DFT) [37] to obtain the thermodynamic bulk equilibrium states of our plate-sphere mixture. The total Helmholtz free energy $F$ is comprised of the ideal and the excess contributions ($F = F^{\rm id} + F^{\rm exc}$). The ideal contribution to $F$ at temperature $T$ for a mixture is given exactly by

$$\beta F^{\rm id} = \sum_i \int dr \int d\boldsymbol{\omega} \rho_i(\boldsymbol{r}, \boldsymbol{\omega})[\ln(\rho_i(\boldsymbol{r}, \boldsymbol{\omega})\Lambda_i^3) - 1], \quad {\rm (A1)}$$

where the sum runs over both species, $\Lambda_i$ is the thermal wavelength of species $i = {\rm p, s}$, and $\rho_i(\boldsymbol{r}, \boldsymbol{\omega})$ is the one-body density profile of species $i$ at position $\boldsymbol{r}$ and orientation specified by the unit vector $\boldsymbol{\omega}$. Since we consider only phases without positional order, we average out all positions $\boldsymbol{r}$ and introduce the angular distribution function $\psi_i(\boldsymbol{\omega})$ of species $i$ via $\rho_i(\boldsymbol{r}, \boldsymbol{\omega}) = \rho_i \psi_i(\boldsymbol{\omega})$ and normalization $\int d\boldsymbol{\omega} \psi_i(\boldsymbol{\omega}) = 1$. For spheres the general formalism simplifies by using a uniform angular distribution function $\psi_{\rm s} = 1/(4\pi)$. Hence, $\rho_i$ is the number density of species $i$ and we obtain

$$\frac{\beta F^{\rm id}}{N} = \sum_i x_i \int d\boldsymbol{\omega} \psi_i(\boldsymbol{\omega})[\ln(\psi_i(\boldsymbol{\omega})\rho_i\Lambda_i^3) - 1], \quad {\rm (A2)}$$

where $N$ is the total number of particles in the system and $x_i$ is the composition of species $i$.

For the excess (over ideal) contribution to the free energy $F^{\rm exc}$, we use an Onsager-like approximation with Parsons-Lee [60,61] rescaling:

$$\frac{\beta F^{\rm exc}}{N} = \Psi(\eta) \rho \sum_{i,j} x_i x_j \int d\boldsymbol{\omega} \int d\boldsymbol{\omega}'$$
$$\times \psi_i(\boldsymbol{\omega}) \psi_j(\boldsymbol{\omega}') V_{i,j}^{\rm ex}(\boldsymbol{\omega}, \boldsymbol{\omega}'), \quad {\rm (A3)}$$

with total density $\rho = \sum_i \rho_i$ and $V_{i,j}^{\rm ex}(\boldsymbol{\omega}, \boldsymbol{\omega}')$ being the excluded volume (i.e., the volume inaccessible to one particle due to the presence of another particle) between particles of species $i$ and $j$ with orientations $\boldsymbol{\omega}$ and $\boldsymbol{\omega}'$, respectively. Here $\Psi(\eta)$ is a scalar function of the total packing fraction $\eta = \rho \sum_i x_i v_i = \sum_i \eta_i$, with $v_i$ being the particle volume of species $i$.

The plates are modelled using hard cylinders (see Fig. 1). For all types of interparticle interactions, (i.e., cylinder-cylinder, cylinder-sphere, and sphere-sphere) there exist analytical expressions for the excluded volumes [17,36,62]. The excluded volume between a cylinder and a sphere, and that between two spheres are independent of the orientation of the particles.

We restrict ourselves to uniaxial situations as reported in the experimental study [11]. Also, in the closely related system of hard cut spheres, the uniaxial nematic is the only stable nematic phase for thickness-to-diameter aspect ratios $\leqslant 0.1$ [63] (note that in our system the aspect ratio is 0.001). Only for aspect ratios $\geqslant 0.15$ the cubatic phase was found to be more stable than the uniaxial nematic [63]. Hence, the angular distribution function of plates depends only on the polar angle $\theta$ (measured with respect to the director)

$$\psi_p(\theta) = \frac{1}{2\pi} \int_0^{2\pi} d\varphi \, \psi_p(\boldsymbol{\omega}). \tag{A4}$$

We, therefore, average in advance the excluded volume over the azimuthal angle $\varphi$ and only retain the polar dependency, $V_{i,j}^{ex}(\theta, \theta')$.

The prefactor $\Psi(\eta)$ in front of the excess free energy in Eq. (A3) rescales the second virial coefficient in the original Onsager's expression [36] by the Carnahan-Starling equation of state [64] of a reference system of hard spheres at the same packing fraction

$$\Psi(\eta) = \frac{4 - 3\eta}{8(1-\eta)^2}. \tag{A5}$$

The topology of the bulk phase diagram does not change with the scaling, which serves to improve the agreement of the $I - N$ transition densities compared to computer simulations [65]. Note that in the low density limit we recover the second virial coefficient, like in the original Onsager expression [36], since $\Psi(\eta \to 0) = 1/2$. Onsager-like density functional theories rely on two-body correlations and can fail to describe the bulk if the symmetries of the stable bulk phases are the result of three- and higher-body correlations [66]. This is not the case here since the excluded volume between two cylinders is minimal if both particles are parallel to each other, i.e., like in the uniaxial nematic phase. We therefore expect the functional to properly describe the topology of the bulk phase diagram.

**Minimization of the functional.** We perform a numerical Picard [67] iteration to minimize the total Helmholtz free energy with respect to the discretized angular distribution function of plates $\psi_p(\theta)$ on a one dimensional grid with 160 points.

We calculate the uniaxial order parameter according to

$$S_p = \int d\theta \frac{3\cos^2(\theta) - 1}{2} \psi_p(\theta). \tag{A6}$$

**Bulk Coexistence.** To obtain the bulk phase diagram we use the Gibbs ensemble and numerically minimize the Gibbs free energy per particle

$$g_b = \frac{F}{N} + \frac{P}{\rho}, \tag{A7}$$

where $P$ is the osmotic pressure and $\rho = \rho_p + \rho_s$ is the total number density. For phase coexistence we need mechanical, thermal and chemical equilibrium. The first two conditions are fulfilled in the Gibbs ensemble by construction ($P$ and $T$ are fixed). To find chemical equilibrium we search for a common-tangent construction on $g_b(x_s)$, with $x_s = \rho_s/\rho$ the composition of spheres. Hence, we numerically minimize the Gibbs free energy per particle $g_b$ with respect to the total density $\rho$ and the orientational distribution function of plates $\psi_p(\theta)$ for fixed values of $P$, $T$, and $x_s$, and then search for a common tangent.

**Average colloidal packing fractions in a sample.** To find the colloidal packing fractions along a sedimentation path we work in the grand canonical ensemble since the paths are lines in the plane of chemical potentials. We minimize the grand canonical potential $\Omega$ per unit of volume

$$\frac{\Omega}{V} = \frac{F}{V} - \rho \sum_i \mu_i x_i, \tag{A8}$$

with respect to $\psi_p(\theta)$, $x_i$, and $\rho$ at fixed values of the chemical potentials. We repeat the minimization for each point along the sedimentation path. From the values of the packing fractions $\eta_i = x_i v_i \rho$ at each point along the path we obtain the average packing fractions $\bar{\eta}_i$, $i = p, s$ in the corresponding sample.